\documentclass[12pt,aps]{revtex4-2}

\usepackage{graphicx} 
\usepackage{amsmath,amssymb,amsfonts,latexsym}
\usepackage[colorlinks=true, linkcolor=blue, citecolor=blue]{hyperref}
\usepackage[mathcal]{euscript}
\usepackage{epsfig}
\usepackage[normalem]{ulem}
\usepackage{color}
\usepackage[en-US]{datetime2}

\usepackage{psfrag}
\usepackage{txfonts,pxfonts,wasysym,}
\usepackage{pifont}
\usepackage{tikz}

\begin{document}

\preprint{APS/123-QED}

\title{Linear stability of turbulent channel flow with one-point closure}

\author{P.V. Kashyap, Y. Duguet, O. Dauchot}
\address{Dept of Mech. Enginneering, Indian Institute of Science, Bengaluru, India.}
\address{LISN-CNRS, Universit\'e Paris-Saclay, Orsay France.}
\address{Gulliver, ESPCI-CNRS, Paris Sorbonne Lettres, France.}

\date{\today}

\begin{abstract}
For low enough flow rates, turbulent channel flow displays spatial modulations of large wavelengths. This phenomenon has recently been interpreted as a linear instability of the turbulent flow. We question here the ability of linear stability analysis around the turbulent mean flow to predict the onset and wavelengths of such modulations.  Both the mean flow and the Reynolds stresses are extracted from direct numerical simulation (DNS) in periodic computational domains of different size. The Orr-Sommerfeld-Squire formalism is used here, with the turbulent viscosity either ignored, evaluated from DNS, or modelled using a simple one-point closure model. Independently of the closure model and the domain size, 
the mean turbulent flow is found to be linearly stable,
in marked contrast with the observed behaviour. This suggests that the one-point approach is not sufficient to predict instability, at odds with other turbulent flow cases. For generic wall-bounded shear flows we discuss how the correct models for predicting instability could include fluctuations in a more explicit way.
\end{abstract}

\maketitle

\section{Introduction}
\begin{figure}
\begin{center}
\includegraphics[width=0.75\columnwidth]{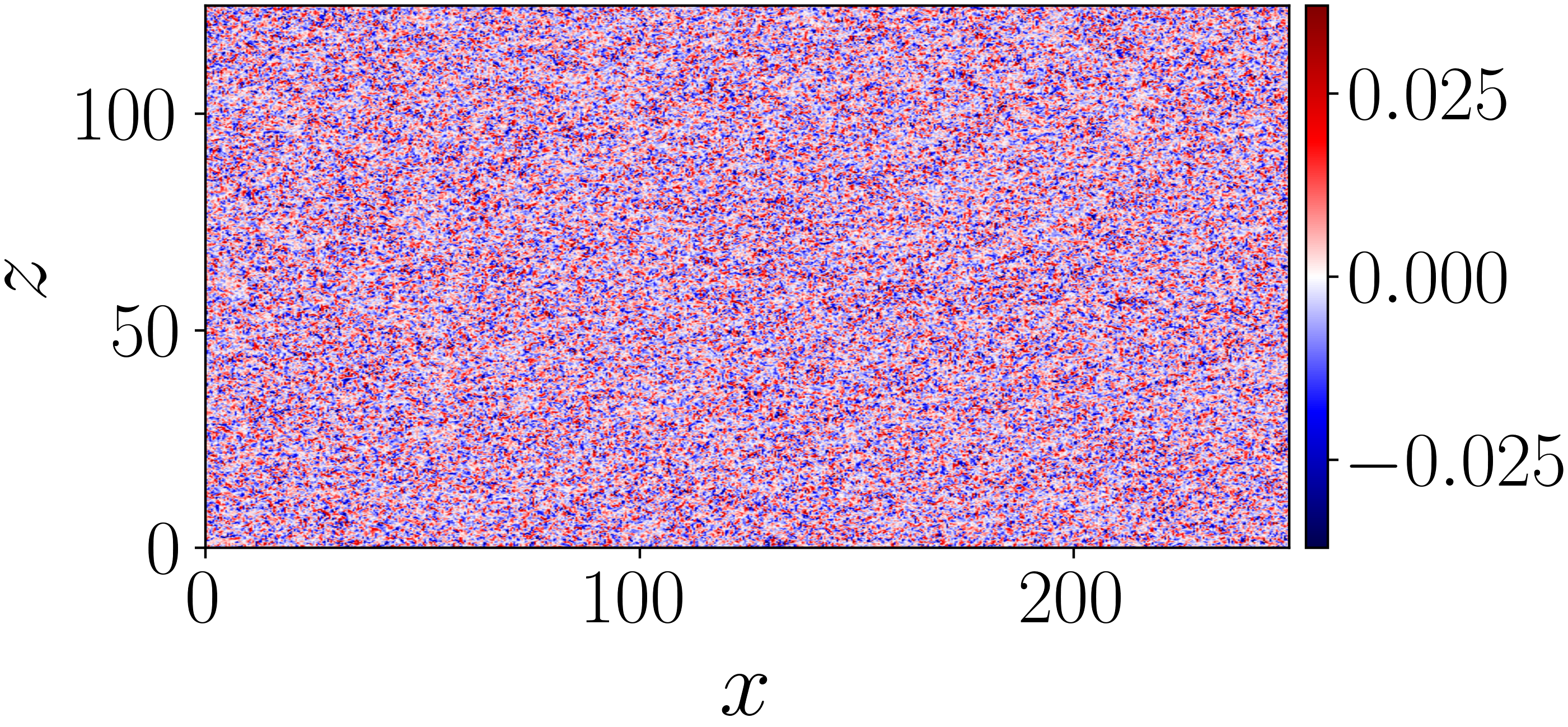}
\includegraphics[width=0.75\columnwidth]{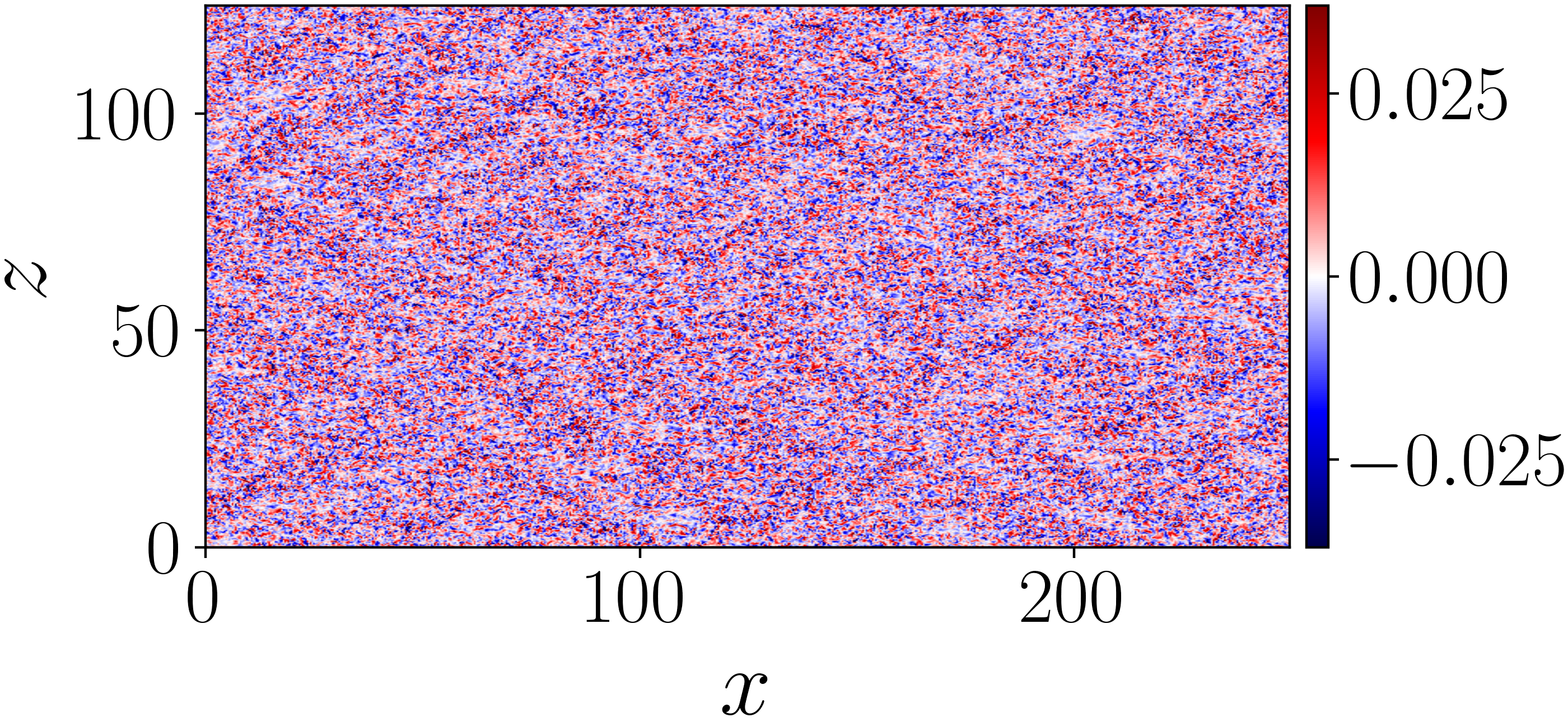}\\
\includegraphics[width=0.75\columnwidth]{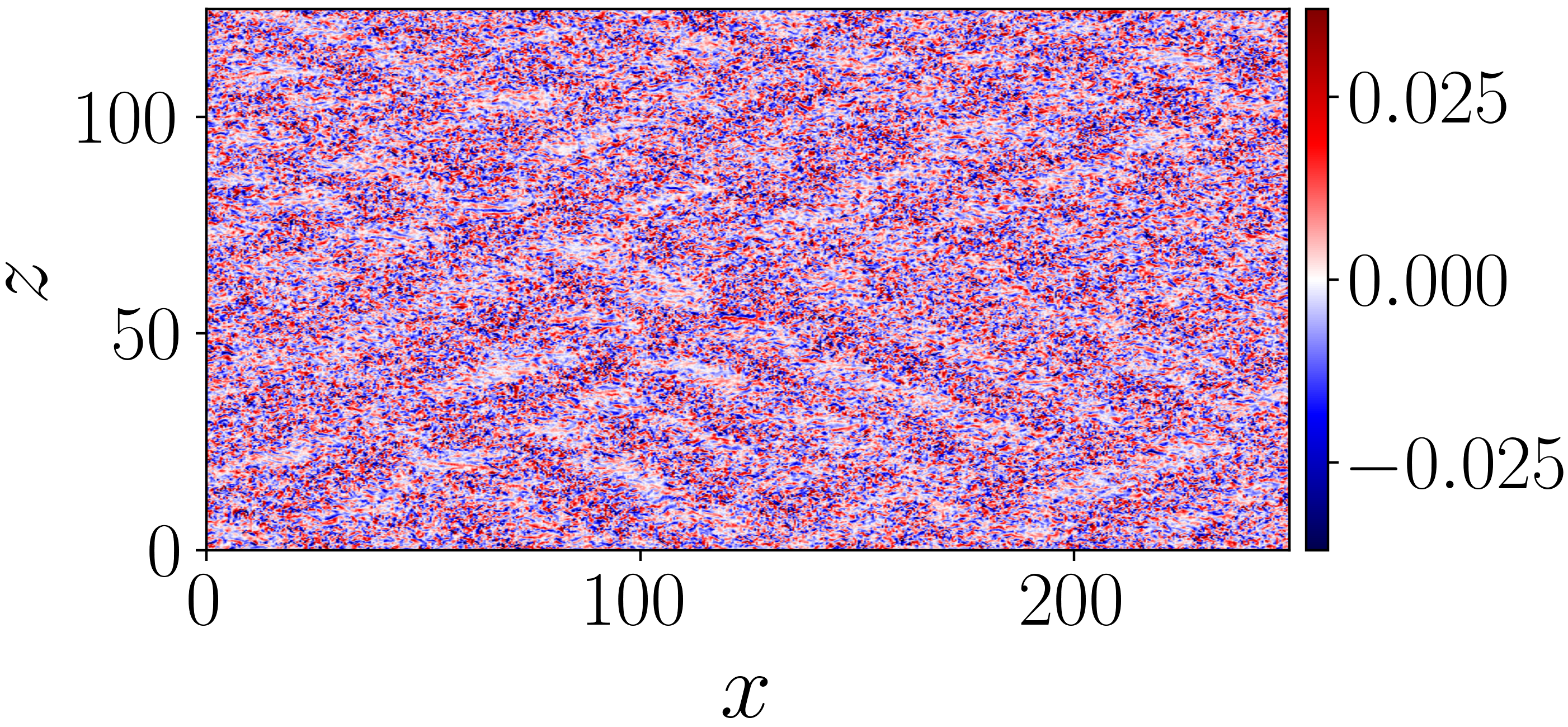}
\includegraphics[width=0.75\columnwidth]{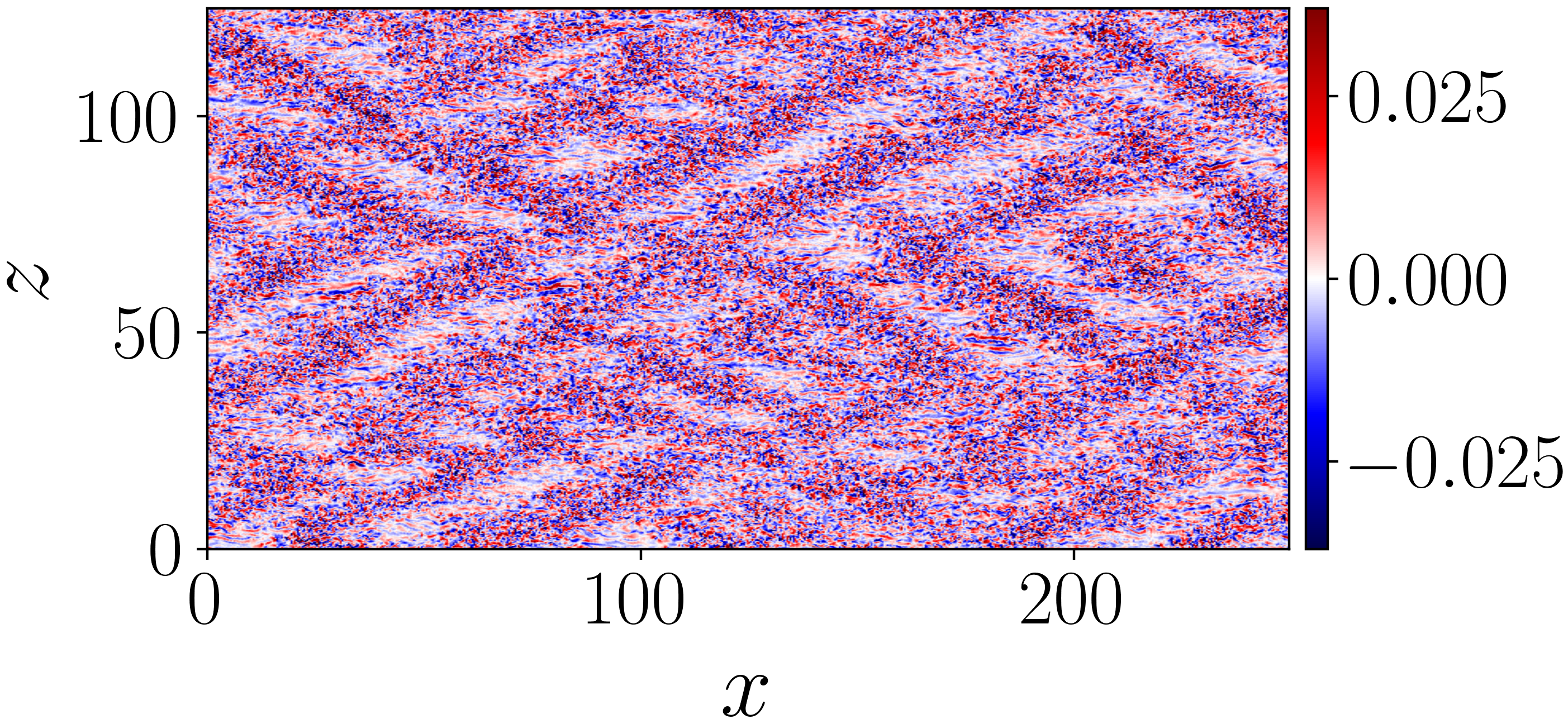}
\end{center}
\vspace{-5mm}
\caption{Onset of laminar-turbulent patterning in turbulent channel flow. Mid-plane (y=0) wall-normal velocity $v(x,z,t)$. From top to bottom : $Re_\tau=100,90,85,80$ in the stationary regime.}
\label{bigpic}
\vspace{-5mm}
\end{figure}

It has been known since the 1960s that many wall-bounded shear flows display a complex organisation akin to laminar-turbulent patterning~\cite{Coles1965transition} at intermediate Reynolds numbers, $Re$. This phenomenon was later reproduced, first experimentally in flows of much larger aspect ratio~\cite{prigent2001spirale,Prigent2002large,Prigent2003long}, then in several numerical studies \cite{tsukahara2005dns,Barkley2005computational,meseguer2009instability,Duguet2010formation}. A large number of recent studies, both experimental and numerical, have described how banded patterns or solitary stripes emerge from the laminar flow after the triggering by localised finite-amplitude disturbances \cite{Duguet2013oblique,tao2018extended,paranjape2019thesis,xiao2020growth,parente2022linear,zhang2023large}. The leading scenario is largely based on the development of a large-scale flow at the interface between laminar and non-laminar zones \cite{Duguet2013oblique}. The route from the turbulent flow to the banded patterns, on the contrary, happens spontaneously and does not require any finite-amplitude perturbation. The emergence of the laminar-turbulent  patterns starting from the fully turbulent regime is illustrated in Fig.~\ref{bigpic} as $Re$ is decreased. The authors of Ref. \cite{Prigent2002large}, suspecting an instability of the spatially homogeneous turbulent flow, modelled the emergence of the resulting patterns using a noisy supercritical Ginzburg-Landau formalism. Later numerical evidence~\cite{tuckerman2020patterns,shimizu2019bifurcations,kashyap2020flow} demonstrated that the precursors of patterning were to be found in the form of low-amplitude modulations of the homogeneous turbulent flow, which is known from simulations at higher $Re$. Finally, an explicit dispersion relation for the underlying instability was reconstructed from costly numerical data, using a response function procedure, and this spatial modulation could be interpreted as a linear instability of the turbulent flow \emph{as a whole}, i.e. including all its fluctuations~\cite{kashyap2022linear}.

It is natural to wonder whether the onset of modulation can be predicted from simpler concepts, such as linear stability, and at a much lower cost. However, linearization about an unsteady turbulent flow is fraught with mathematical difficulties if all temporal and spatial fluctuations are taken into account. It would thus be desirable to average out the fluctuations and perform the linear stability analysis of the mean flow. However, because of the non-linearity of the original equations, that for the mean flow includes correlations of the fluctuations and calls for \emph{turbulent closures}. 

Following this line of thought, it was conjectured that turbulent mean flows are marginally stable~\cite{malkus1956outline}, an idea furthermore suggested as a governing principle used to predict mean flows. Focusing on the canonical case of channel flow between two infinite plates, driven by a pressure gradient, this hypothesis was later dismissed in favor of strict stability \cite{reynolds1967stability,reynolds1972mechanics},  in the sense that all eigenvalues of the linearized problem are strictly negative. However, the range of values of $Re$ considered in these studies was always too high to be relevant to the pattern formation problem of interest here. The present paper is devoted to a similar stability analysis of turbulent channel flow restricted to the range of $Re$ values where laminar-turbulent patterns are known to occur.

There has been a long debate as to whether the linear stability or instability of the mean turbulent flow could be considered as a reliable prediction \cite{reynolds1972mechanics,beneddine2016conditions}. It fails in some cases when the spectral distribution of energy of the mean flow is not steep enough~\cite{sipp2007global}, but it proved accurate in other situations, such as the two-dimensional cylinder wake~\cite{barkley2006linear}, in rotating channel flow \cite{brethouwer2014recurrent}, thermo-solutal convection~\cite{turton2015prediction}, and more recently in turbulent convection~\cite{cossu2022onset} and magneto-hydrodynamic turbulent flows ~\cite{ponty2007dynamo}. Furthermore, finite-time stability~\cite{Schmid2007nonmodal} of the turbulent mean flow indicates strong transient growth, at spanwise scales consistent with the appearance of both near-wall streaks and large-scale streaks \cite{del2006linear,pujals2009note}. This encouraging result suggests that linearization about a mean flow can contain decisive information about the fluctuation fields and be a predicting tool.

The wall turbulence community has possibly ignored the possibility that such instabilities of the turbulent regime could occur at low rather than at high $Re$. Nevertheless, linearizing around a well-defined mean flow, as formally done at high Reynolds number, remains \emph{a priori} a relevant approach to capture the emergence of the long wavelength modulation observed in shear flows when \emph{decreasing} the Reynolds number. As a further motivation to consider the instability formalism, a recent low-order model for the mean flow of a companion shear flow (the so-called Waleffe flow) has been recently suggested \cite{benavides2023model}, and the fixed point associated with the mean turbulent flow displays indeed a linear instability precisely at the onset of the turbulent bands.

In this work we revisit the approach using the Orr-Sommerfeld-Squire formalism applied to this problem by focusing on the relevant parameter range. Both the mean flow and the turbulent viscosity are extracted from direct numerical simulation in periodic computational domains of varying size. As an alternative the turbulent viscosity can also be modelled using a simple one-point closure model. The mean turbulent flow is found to be linearly stable, regardless of the choice of one-point model and domain size, in marked contrast with the observed behaviour. We conclude that one-point closure models associated with one-dimensional base flows do not capture the essence of the instability mechanism and discuss how the proper fluctuations could be included in a more explicit way.
 
\section{Closure formalism}
\subsection{Instability formalism for mean flows}

We start from the original momentum equations in vector form,
\begin{eqnarray}
\frac{\partial {\bm u}}{\partial t} + ({\bm u} \cdot \nabla){\bm u}= -\frac{1}{\rho}\nabla p + \nabla \cdot (\nu \nabla {\bm u}),
\label{eq:u}
\end{eqnarray}
which  together with the incompressibility condition $\nabla \cdot {\bm u}=0$ forms the Navier-Stokes equations. In Eq. \ref{eq:u}, ${\bm u}=(u_x,u_y,u_z)$ is the (dimensional) three-dimensional velocity field, $p$ the pressure field, $\nu$ the kinematic viscosity and $\rho$ the fluid density. $x$ denotes the streamwise coordinate, $y$ the wall-normal coordinate where the walls are at $y=\pm h$, and $z$ is the spanwise coordinate. A dimensionless Reynolds number $Re_{\tau}=u_{\tau}h/\nu$ can be built based on the friction velocity $u_{\tau}$, related to the imposed mean wall-shear stress $\tau$ by the relation $\rho u_{\tau}^2=\tau$.

Any averaging $\overline{(\cdot)}$ satisfying Reynold's conditions \cite{monin1971statistical} is admissible here, such that \emph{a priori} the average ${\bm U}=\overline{{\bm u}}$ of a velocity field ${\bm u}(x,y,z,t)$ can also depend on $x$,$y$,$z$ and $t$.
By applying the chosen averaging to all terms of the original equation~\ref{eq:u} one obtains
\begin{eqnarray}
\frac{\partial {\bm U}}{\partial t} + 
({\bm U} \cdot \nabla){\bm U}= -\nabla P + \nabla \cdot (\nu \nabla {\bm U}) - \nabla \cdot (\overline{{\bm u'}\otimes{\bm u'}})
\label{eq:URS}
\end{eqnarray}
where ${\bm u'} = {\bm u} - {\bm U}$, and the term $\overline{{\bm u'}\otimes{\bm u'}}$, forms the Reynolds stress tensor.
%
This mean flow equation for ${\bm U}$ is not closed. The standard strategy amounts to deriving an evolution equation for the Reynolds stress, which in turn will depend on higher order correlations. To make progress, the resulting hierarchy of equations must then be truncated with a closure model. 

When the flow geometry is simple enough -- little or no anisotropy, zero or small curvature, ... --, as is the case here, a reasonable assumption, the Boussinesq hypothesis, consists in assuming that the Reynolds stress anisotropy tensor is aligned with the strain tensor of the mean flow. 
The proportionality coefficient $\nu_t$ is called the turbulent eddy viscosity. An explicit expression for $\nu_t$ is called a one-point closure model, in contrast to closure models based on the correlations between several spatial locations.
Equation~\ref{eq:URS} can then formally be re-written in the more compact form 
\begin{eqnarray}
\frac{\partial {\bm U}}{\partial t} + 
({\bm U} \cdot \nabla){\bm U}= -\nabla P + \nabla \cdot (\nu_T \nabla {\bm U}) 
\label{eq:UNUT}
\end{eqnarray}
where $\nu_T=\nu+\nu_t$.\\

The mean flow equation has a particular steady solution ${\bm U^*}$ corresponding to the stationary turbulent flow found at high $Re$ (and different from the laminar Poiseuille solution). The channel flow geometry is furthermore invariant by translation in the streamwise and spanwise directions. In such a case, a homogeneous and steady turbulent flow is observed : all turbulent statistics of this flow, in particular the mean flow and the turbulent viscosity, become independent of $t$, $x$ and $z$.  The turbulent viscosity can be evaluated as

\begin{eqnarray}
\nu_t=- 
\frac{
\overline{u'_xu'_y}}{\frac{d}{dy}U_x}.
\label{eq:RS}
\end{eqnarray}

Because equation~\ref{eq:UNUT} is also a nonlinear PDE, there is no reason why ${\bm U^*}$ should be the only stationary solution. More specifically, the symmetries associated with the translation invariance in the streamwise and spanwise directions can be spontaneously broken following a classical instability scenario. In the simplest case, such a scenario would start by the linear destabilization of the solution ${\bm U^*}$, which we now investigate by applying the Orr-Sommerfeld-Squire formalism to the \emph{mean} flow equation~\ref{eq:UNUT}.

\subsection{Orr-Sommerfeld-Squire system}

Assuming a small perturbation $\tilde{{\bm U}}(x,y,z,t)$ added to ${\bm U^*}(y)$, it obeys the linearized equation
\begin{eqnarray}
\frac{\partial \tilde{\bm U}}{\partial t} + (\tilde{\bm U} \cdot \nabla){\bm U^*} +  ({\bm U^*} \cdot \nabla)\tilde{{\bm U}}= -\nabla \tilde{P} + \nabla \cdot (\nu_T \nabla \tilde{\bm U}) 
\label{eq:UNUT2}
\end{eqnarray}
The linear stability analysis follows closely the classical Orr-Sommerfeld theory (see e.g. \cite{schmid2002stability}), except that (i) the laminar base flow ${\bm U}_{lam}(y)$ is replaced by the mean base flow ${\bm U^*(y)}$, and (ii) the viscosity $\nu_T$ depends on $y$ \cite{morra2019relevance}. Eq. \eqref{eq:UNUT2} can also be obtained from the triple decomposition suggested in Ref. \cite{reynolds1972mechanics}, where the total flow is decomposed into a mean flow, a coherent wave part and less organised turbulent fluctuations.

Because of the spatial and temporal translation symmetries of Eq. \ref{eq:UNUT}, we use the classical velocity ansatz $ \tilde{{\bm U}} \sim \hat{{\bm U}}(y)e^{i(\alpha x + \beta z - \omega t)}$, where $\alpha$ and $\beta$ are respectively the axial and spanwise wavenumber, and $\omega=\omega_r+i\omega_i$ a complex number, with $\omega_r$ the angular frequency and $\omega_i$ the growth rate of the perturbations. Expanding all scalar fields on a basis of $N$ Chebyshev polynomials ($N=65$, throughout the study), the original system takes, for each choice of $\alpha$ and $\beta$, the form of a generalised matrix eigenvalue problem \cite{schmid2002stability,schmid2014analysis} ${\bm A}{\bm q}=-i\omega {\bm q}$ where 
\begin{equation}
{\bm A}=\begin{pmatrix}
\Delta^{-1}\mathcal{L}_{OS} & 0 \\
-i\beta U' & \mathcal{L}_{SQ}
\end{pmatrix}
\label{eq:LinOp}
\end{equation}
with
\begin{eqnarray}
   \mathcal{L}_{OS}=-i\alpha(U\Delta-U'')+ \nu_T\Delta^2 + 2\nu'_T\Delta\mathcal{D} + \nu''_T(\mathcal{D}^2+k^2)+ \nu_T\Delta+ \nu'_T\mathcal{D}
   \label{eq:LOS}
\end{eqnarray}
and
\begin{eqnarray}
   \mathcal{L}_{SQ}=-i\alpha U + \nu_T\Delta+ \nu'_T\mathcal{D}.
   \label{eq:LSQ}
\end{eqnarray}

The vector ${\bm q}$ contains the wall-normal velocity $\hat{U}_y$ and the wall-normal vorticity $\hat{\omega}_y$, respectively, expressed at all (discrete) collocation points in $y$. The differential operators are $\mathcal{D}=d/dy$ (just like the prime symbol) and $\Delta:=\mathcal{D}^2-k^2$, with $k^2=\alpha^2+\beta^2$. The formulation is identical to that in Ref. \cite{morra2019relevance}.

\subsection{Teachings from the Squire theorem} \label{sec:squire}

Since the formalism is mathematically similar to the Orr-Sommerfeld-Squire, a legitimate question is the applicability and the consequences of the Squire transformation. As a consequence of this transformation, if there exists a solution of the Orr-Sommerfeld eigenvalue problem its growth rate satisfies the scaling $\omega_i(\alpha,\beta,Re) = \alpha\, c_i(k^2,\alpha Re)$, where $c_i$ is the imaginary part of a complex phase velocity. Accordingly for any solution parameterised by $(\alpha_1,\beta_1,{\omega_i}_1, R_1)$, there exists an identical solution with $k_2^2 = k_1^2$, ${\omega_i}_2 = \alpha_2 {\omega_i}_1 / \alpha_1$ and $R_2 = \alpha_1 R_1 / \alpha_2$. This is usually used to prove that if there is a three dimensional linear instability of a laminar plane shear flow (${\omega_i}_1=0, \alpha_1>0, \beta_1>0$) for a given critical Reynolds number ${Re_c}$, then there is a two-dimensional one (${\omega_i}_2=0, \alpha_2>\alpha_1, \beta_2=0$) emerging at a lower critical Reynolds number~\cite{Drazin2004hydrodynamic}.But it also says that there are other marginally stable modes at higher Reynolds number (${\omega_i}_2=0, \alpha_2<\alpha_1, \beta_2>\beta_1$).
n the usual case, the instability sets in when increasing the Reynolds number and this second part of the theorem is useless, since the base flow is already unstable. Here, in contrast, we are looking for an instability taking place when decreasing $Re$. Using the theorem, one would conclude that there are unstable modes at all Reynolds numbers (${\omega_i}_2=0, \alpha_2 \rightarrow 0, \beta_2>\beta_1,  R_2\rightarrow \infty$). This is certainly not backed by experimental or numerical evidence. \\

Note that the scaling leading to the Squire transformation does not hold strictly, because the base flow and the turbulent viscosity depend on the Reynolds number, in contrast with the laminar flow, an important property of plane shear flows. Nevertheless, close enough to the three dimensional critical value of Re, one does not expect any rapid change of the base flow and a weaker formulation of the theorem should hold. In such an scenario, given that all observations point towards a well defined three-dimensional pattern, this suggests that the Orr-Sommerfeld-Squire formulation, or equivalently any linear stability analysis of the mean flow alone, is likely to fail at predicting the observed instability where both critical $\alpha$ and $\beta$ are non-zero. In the present work, we will actually show explicitly that this is indeed the case for the plane channel flow.\\

\subsection{Models for the turbulent viscosity}
The Boussinesq eddy viscosity model itself does not prescribe the form of the turbulent viscosity $\nu_t(y)$, which depends on the flow geometry and the Reynolds number. We consider here three different simple explicit models for $\nu_t$.

The most drastic one trivially assumes vanishing Reynolds stresses : $\nu_t=0$. It amounts to the wishful hypothesis that linearization around the mean flow could be carried out as if it were a solution to the Navier-Stokes equations. This pathological situation is a special case of the Cess formula with $\kappa=0$ (or also $A \rightarrow \infty$). Counter-intuitively perhaps, this procedure leads in some flow cases to physically correct results~\cite{barkley2006linear,sipp2007global}, and it should hence not be immediately discarded. 

Second, we use an analytical closure model, popular for high-Reynolds-number channel flow, proposed by Cess~\cite{cess1958survey,pujals2009note}. The Cess formula reads :
\begin{eqnarray}
    \frac{\nu_{T}}{\nu}=\frac{1}{2}\left( 1 + \frac{\kappa^2 Re_{\tau}^2(1-\eta^2)^2(1+2\eta^2)^2(1-\exp{((|\eta|-1)Re_{\tau}/A)^2}}{9}\right)^{\frac{1}{2}}+ \frac{1}{2}
    \label{eq:Cess}
\end{eqnarray}
where $\eta=y/h$ and with $\kappa$=0.426 and $A$=25.4 as in~\cite{pujals2009note}. The Cess formula has been used also in similar studies of turbulent mean flow stability in e.g. Refs~\cite{reynolds1972mechanics}, \cite{del2006linear}, \cite{morra2019relevance} and~\cite{cossu2022onset}. Note that the previous model with vanishing Reynolds stresses corresponds to a special case of the Cess closure with $\kappa=0$.

Other popular modelling possibilities for turbulent channel using one-point closures are the $k-\varepsilon$ and $k-\Omega$ closures, where $\nu_t$ is imposed as a scalar field built on the mean kinetic energy and dissipation. Linear stability analysis of the $k-\Omega$ model with prescribed constants has been shown to predict strict stability of the mean plane Couette flow~\cite{tuckerman2010instability}. Because of the many constants involved in this model, usually determined from benchmarks performed at high $Re$ rather than in the transitional regime, we did not pursue this approach here.

In the framework of Unsteady RANS modelling (URANS), Closure obtained by the compressible form of the Spalart-Allmaras  turbulence model has also been considered. In particular, in the study of Ref. \cite{crouch2007predicting}, the turbulent viscosity is considered as an unknown fluctuating variable in its own right. Fluctuations of $\nu_t$ are hence also taken into consideration through the use of yet another closure model. However the models used in such studies, for instance the Spalart-Allmaras model in Ref. \cite{crouch2007predicting}, are phenomenological rather than based on any governing principle. This approach has hence not been pursued here.\\

Instead, we propose a more direct approach, in avoiding the formulation of any model and directly estimating $\nu_t(y)$ from direct numerical simulation (DNS), numerically evaluating the Reynolds stresses and the mean flow $U^*_x(y)$, and using Eq.~\ref{eq:RS}. The  apparent singularity at $y=0$, where the denominator vanishes, is avoided by using L'H\^{o}pital's rule. 
Stability around a numerically determined turbulent mean flow is now routinely evaluated e.g. in resolvent analysis studies~\cite{hwang2010amplification}. However, considering the Reynolds stresses under a data-driven approach (as in e.g. Ref. \cite{farano2017optimal}) has, to our knowledge, not been attempted in linear stability studies despite the marginal additional cost. 

\section{Numerical data}
All simulations were performed with the MPI-parallel spectral solver Channelflow2.0 \cite{channelflow2,kashyap2020flow} in a domain of size $L_x \times L_z=250 \times 125$ and averaged over times up to $t=4 \times 10^3$ time units.  They have been performed at the French computation center IDRIS using up to 640 cores. The pseudo-spectral resolution is based on 1024 modes in both $x$ and $z$ (including dealiasing with the 2/3 rule), 65 in $y$, which is comparable to~\cite{shimizu2019bifurcations}.

The averages considered in this study, i.e. the steady mean flow and the Reynolds stresses
\begin{eqnarray}
{\bm U}(y)=\lim_{T \to \infty}\frac{1}{T}\int_0^T{\bm u}(x,y,z,t)dt,\\
\overline{u'_xu'_y}(y)=\lim_{T \to \infty}\frac{1}{T}\int_0^T [(u_x-U_x)u_y](x,y,z,t)dt.
\end{eqnarray}
have been evaluated in practice as time averages and computed over 4,000 time units after excluding transients.\\

\begin{figure}
\begin{center}
\includegraphics[width=0.7\columnwidth]{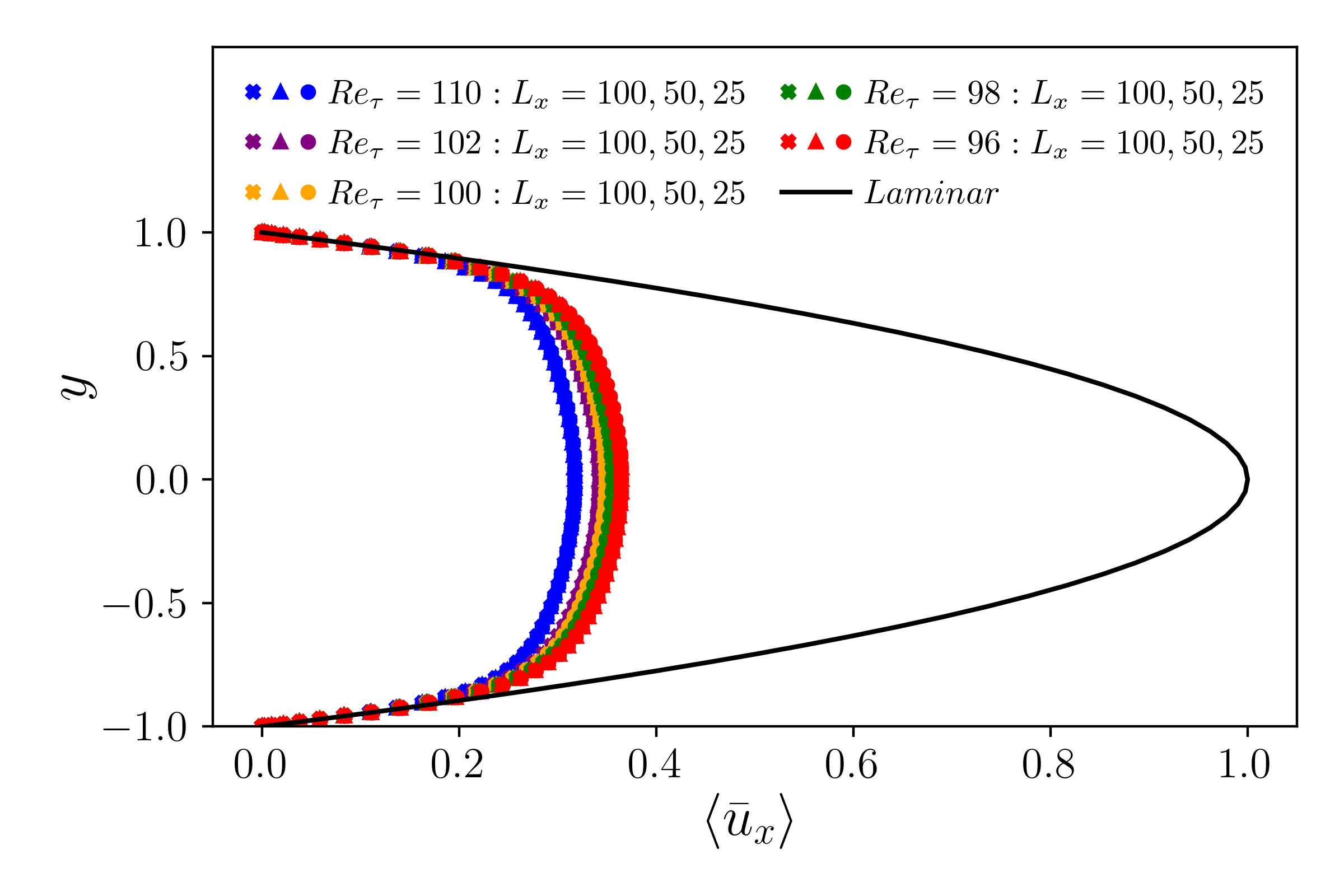}
\end{center}
\vspace{-5mm}
\caption{Mean flow $U^*_x(y)$ from DNS for $Re_{\tau}$ from 110 to 96, used for the linear stability analysis. Different domain sizes have been used without any strong influence on the statistics.}
\label{fig:meanflows}
\vspace{-0mm}
\end{figure}
\begin{figure}[b!]
\begin{center}
\includegraphics[width=0.49\columnwidth]{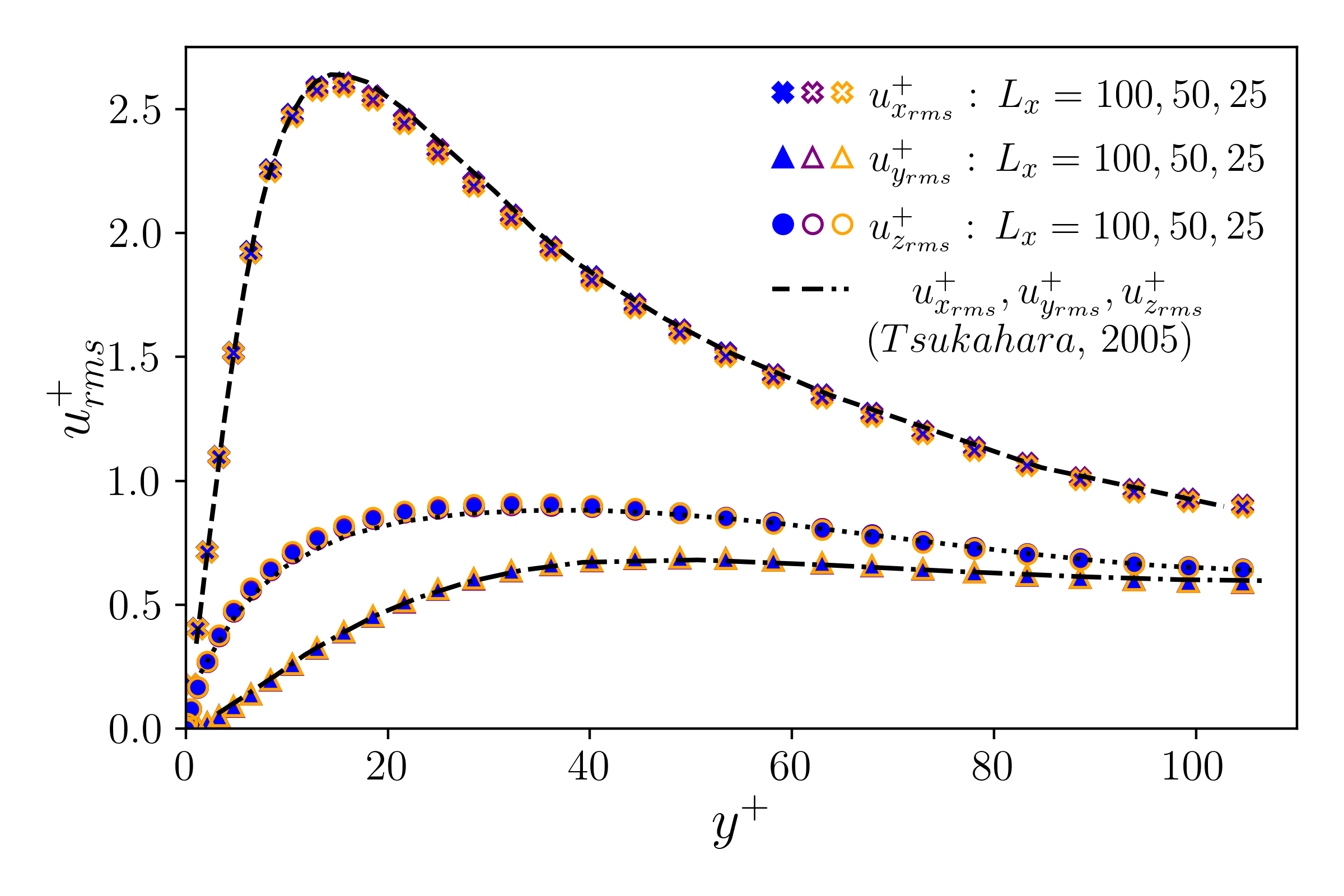}
\includegraphics[width=0.49\columnwidth]{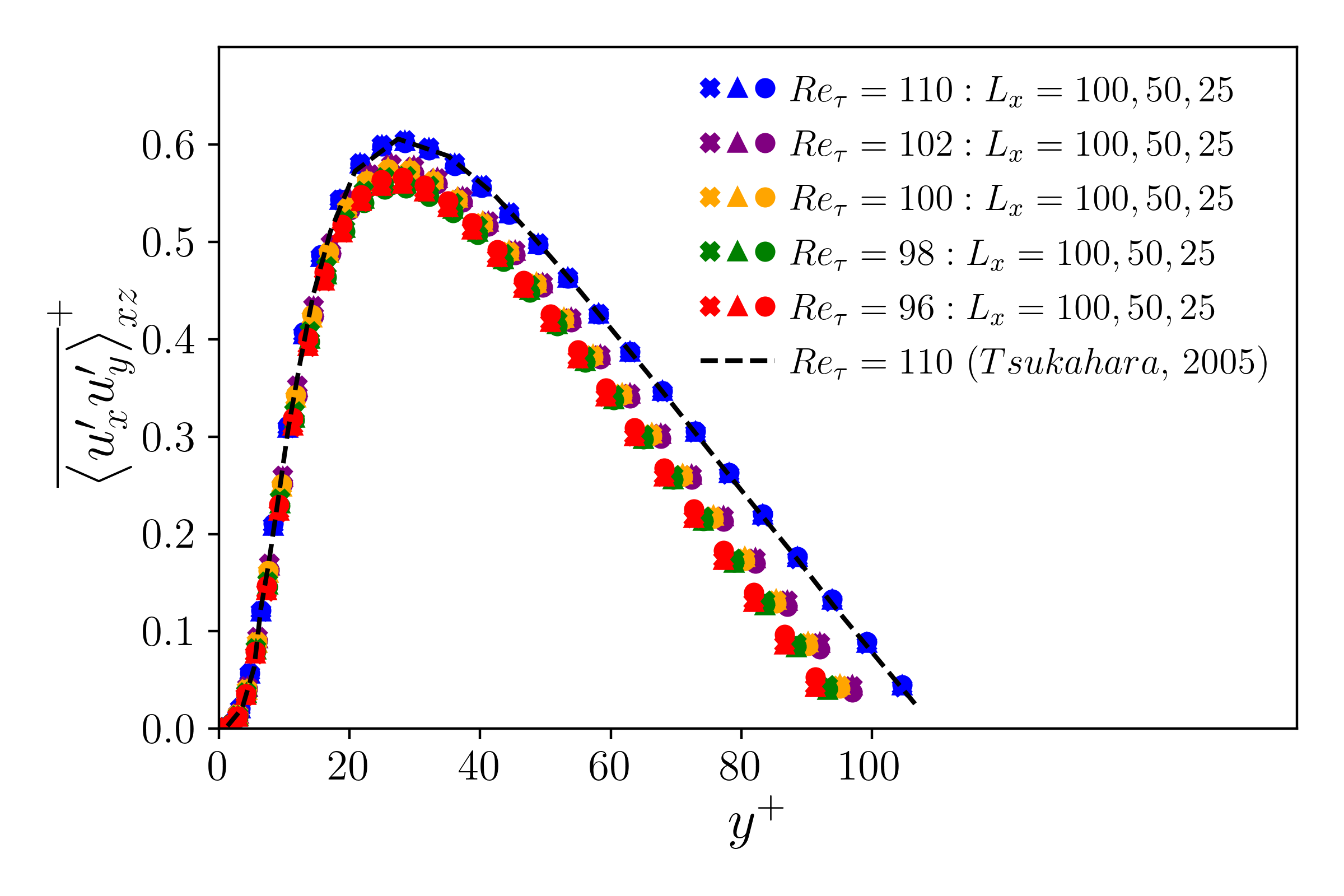}
\end{center}
\vspace{-5mm}
\caption{R.M.S. streamwise velocity fluctuations for $Re_{\tau}=110$ (Left) and Reynolds stress $\overline{<u'_xu'_y>}$ for $Re_{\tau}$ from 110 to 96 (Right), versus $y^+$. Includes a comparison to \cite{tsukahara2005dns}.}
\label{fig:RMS}
\vspace{-0mm}
\end{figure}

Different computational domains have been tested for this study, respectively  $L_x=2L_z=4\pi,25,50$ and $100$. Note that the computational domain size is not directly related to the wavenumbers $\alpha$ and $\beta$ emerging from stability analysis.
For $L_x=4\pi$, the Reynolds stresses are not symmetric in $y$ over moderate observation times, and statistics are slow to converge. This asymmetry in $y$ of the flow fields in computational domains that are too small has long been documented \cite{Jimenez1991minimal} and we discard the corresponding data.  For the larger system sizes $L_x=25,50,100$, the results are quantitatively similar as demonstrated, for instance, by the mean flow profiles $U_x^*(y)$, computed by DNS for different values of $Re_{\tau}$ from 110 down to 96 and for different domain sizes that are compared in figure~\ref{fig:meanflows}. All results displayed in the following were obtained with a system size $L_x=100$. We also checked that the fluctuations extracted from the present DNS simulations match those in the literature, see Figure~\ref{fig:RMS}.

\begin{figure}
\begin{center}
\includegraphics[width=0.6\columnwidth]{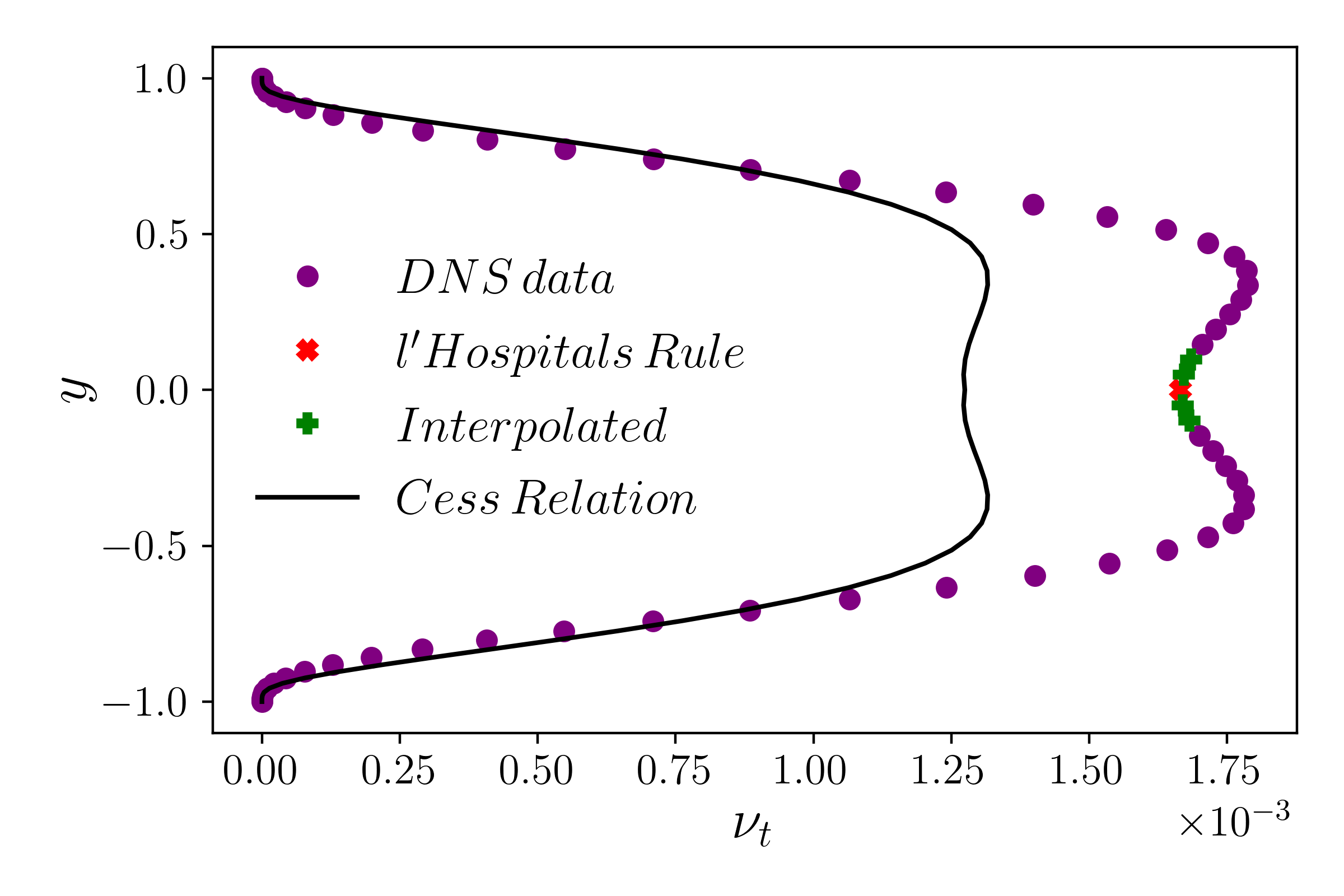}
\end{center}
\vspace{-5mm}
\caption{Turbulent viscosity $\nu_t$ for $Re_\tau=102$, evaluated directly from DNS using Eq. \ref{eq:RS} (dots), interpolated near the midplane $y=0$ using l'Hospitals rule, and using the Cess formula of Eq. \ref{eq:Cess} (line).}
\label{fig:nutcess}
\vspace{-5mm}
\end{figure}

Finally Figure~\ref{fig:nutcess} compares the turbulent viscosity profiles, evaluated directly from the DNS simulations using Eq. \ref{eq:RS} or from the Cess formula, Eq.~\ref{eq:Cess}. Not so surprisingly, the Cess formula, designed for high Reynolds flows, underestimates the turbulent viscosity.

\section{Stability results}


\begin{figure}
\begin{center}
\includegraphics[width=0.46\columnwidth]{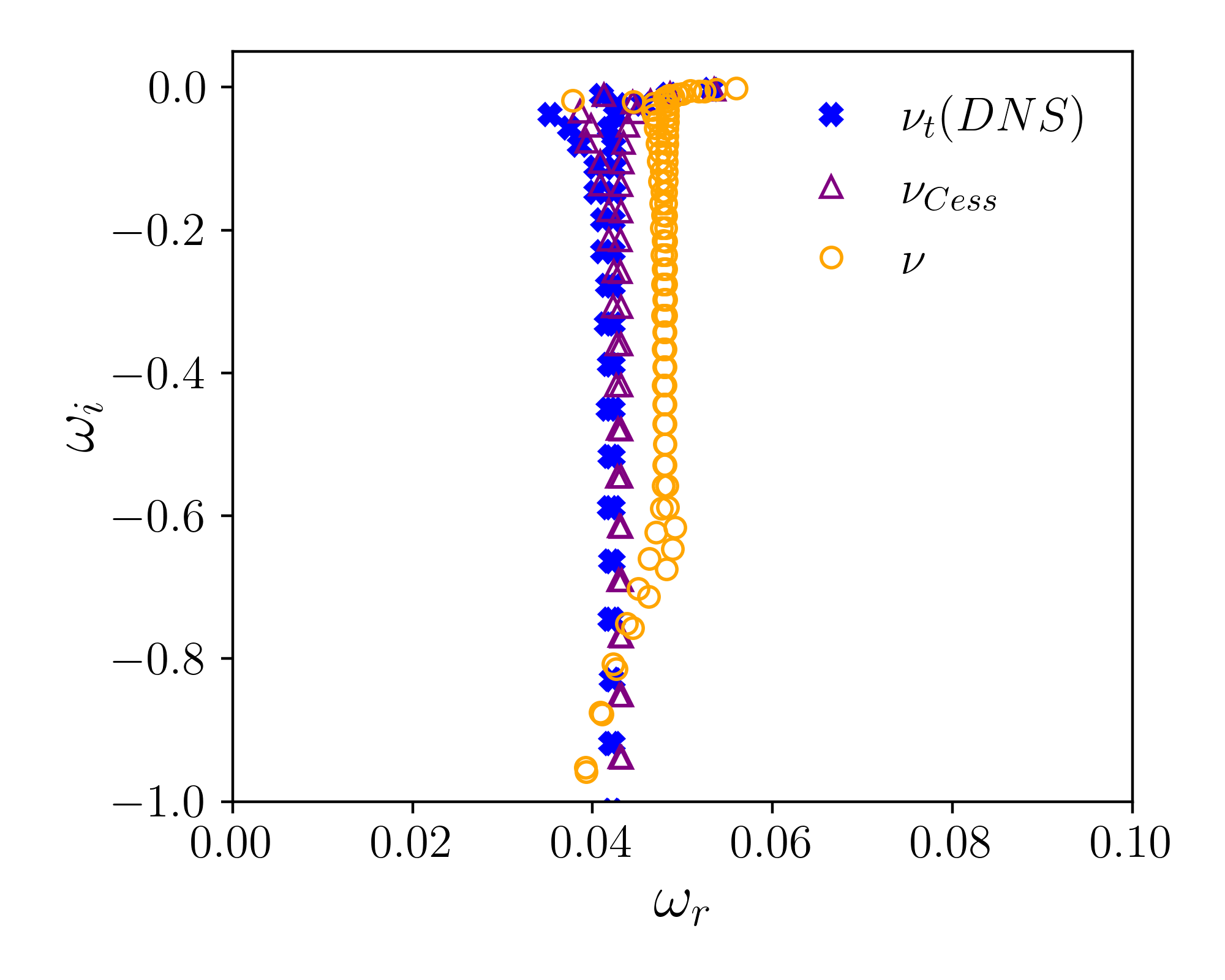}
\includegraphics[width=0.46\columnwidth]{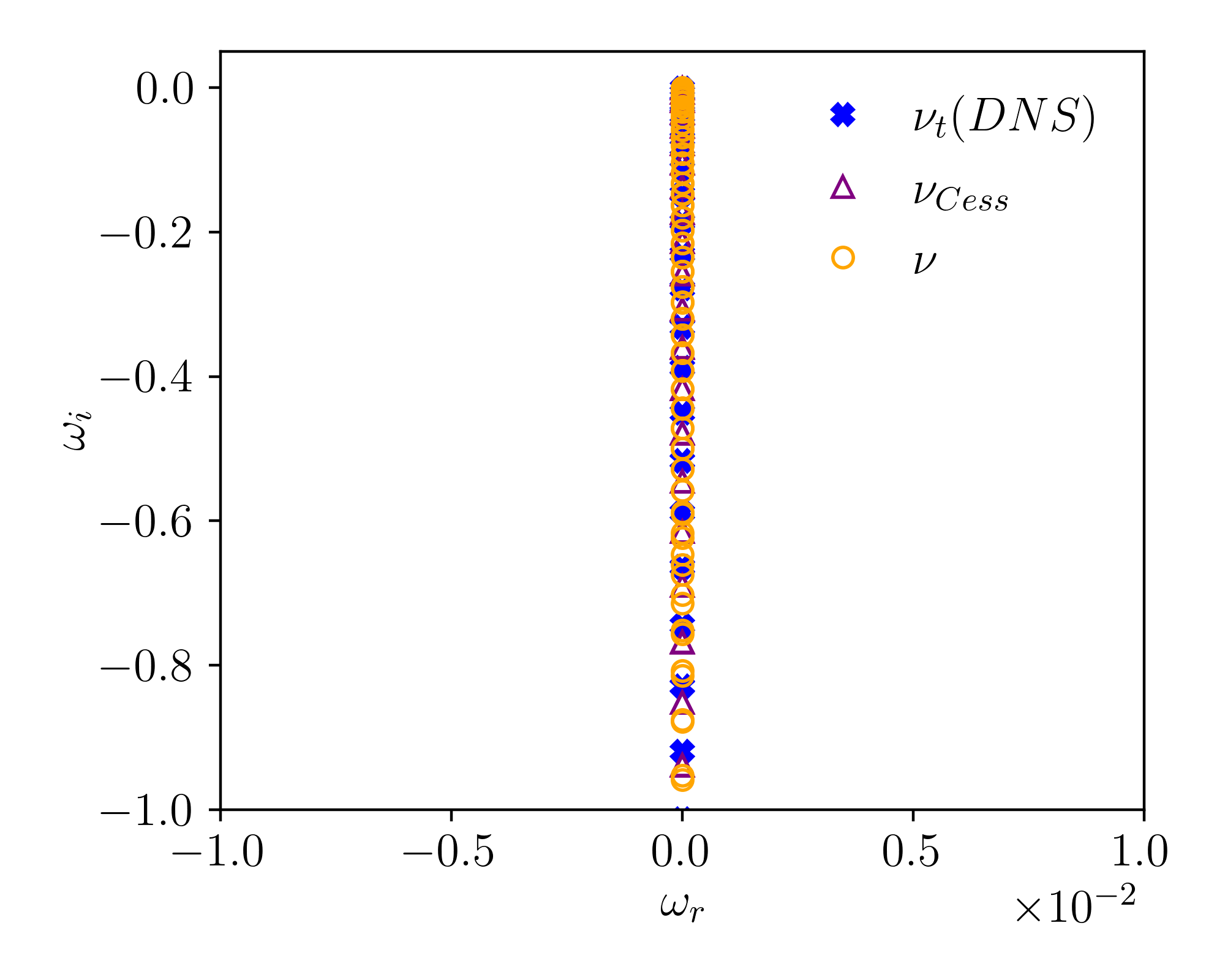}\\
\includegraphics[width=0.46\columnwidth]{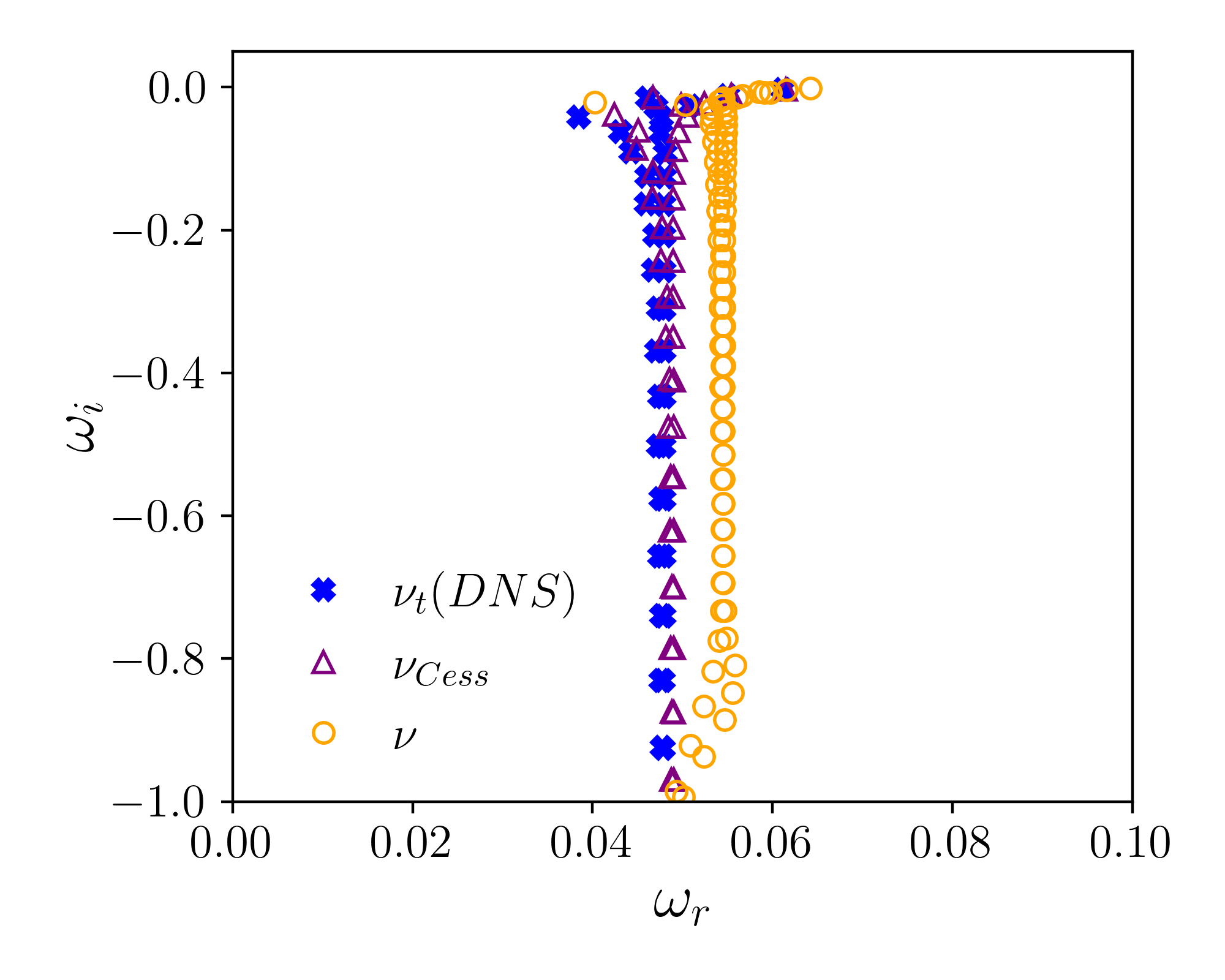}
\includegraphics[width=0.46\columnwidth]{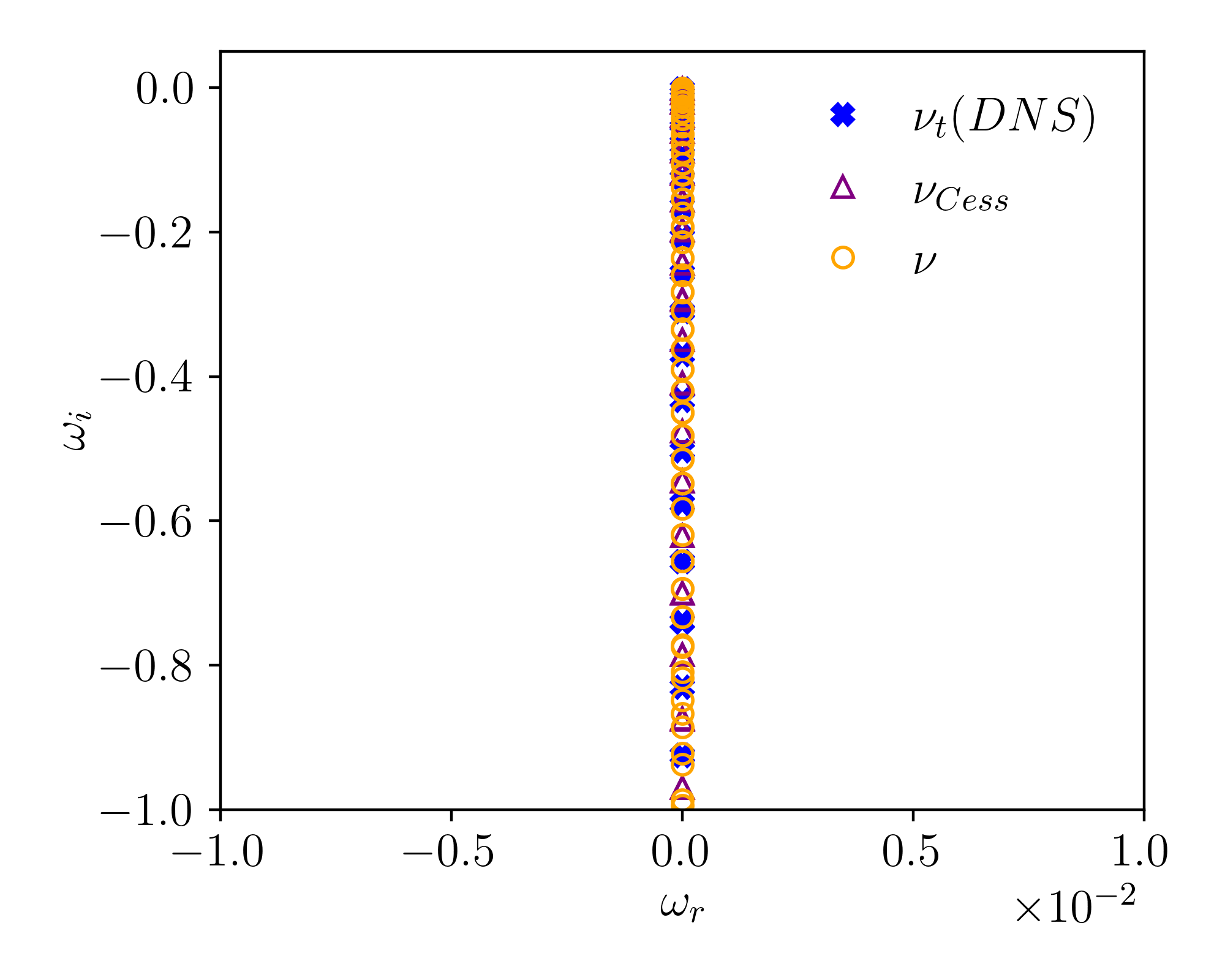}\\
\end{center}
\vspace{-5mm}
\caption{Complex eigenvalues in the $\omega$-plane for $Re_\tau=110$ (top row) and $Re_\tau=96$ (bottom row). Left column : $(\alpha,\beta)=(0.18,0.42)$ (the critical wavevector in Ref. \cite{kashyap2022linear}), Right column : $(\alpha,\beta)=(0,0)$.}
\label{fig:decayrates}
\vspace{-0mm}
\end{figure}

The eigenvalues of the linear operator defined by equations~\ref{eq:LinOp},~\ref{eq:LOS} and~\ref{eq:LSQ} have been computed using Python for values of $\alpha$ and $\beta$ on a discrete grid. To illustrate the results, we have picked two specific values of $(\alpha,\beta)$, respectively $(0.18,0.42)$, the marginal wave-vector expected from the response analysis conducted in~\cite{kashyap2022linear}, expressed in units of $2\pi/h$, and $(0,0)$. Figure~\ref{fig:decayrates} displays the growth rate of the perturbation for two values of $Re_{\tau}$ larger than the expected instability threshold $Re_{\tau}^c$, estimated in~\cite{kashyap2022linear}. In all cases the growth rates are strictly negative except for two neutral modes corresponding to the neutral directions $x$ and $z$ : perturbations to the mean relax towards the linearly stable mean flow $U^*$. For $(\alpha,\beta)=(0,0)$ all modes are stationary and decaying. For $(\alpha,\beta)= (0.18,0.42)$ the modes are propagating. 

\begin{figure}
\begin{center}
\includegraphics[width=0.46\columnwidth]{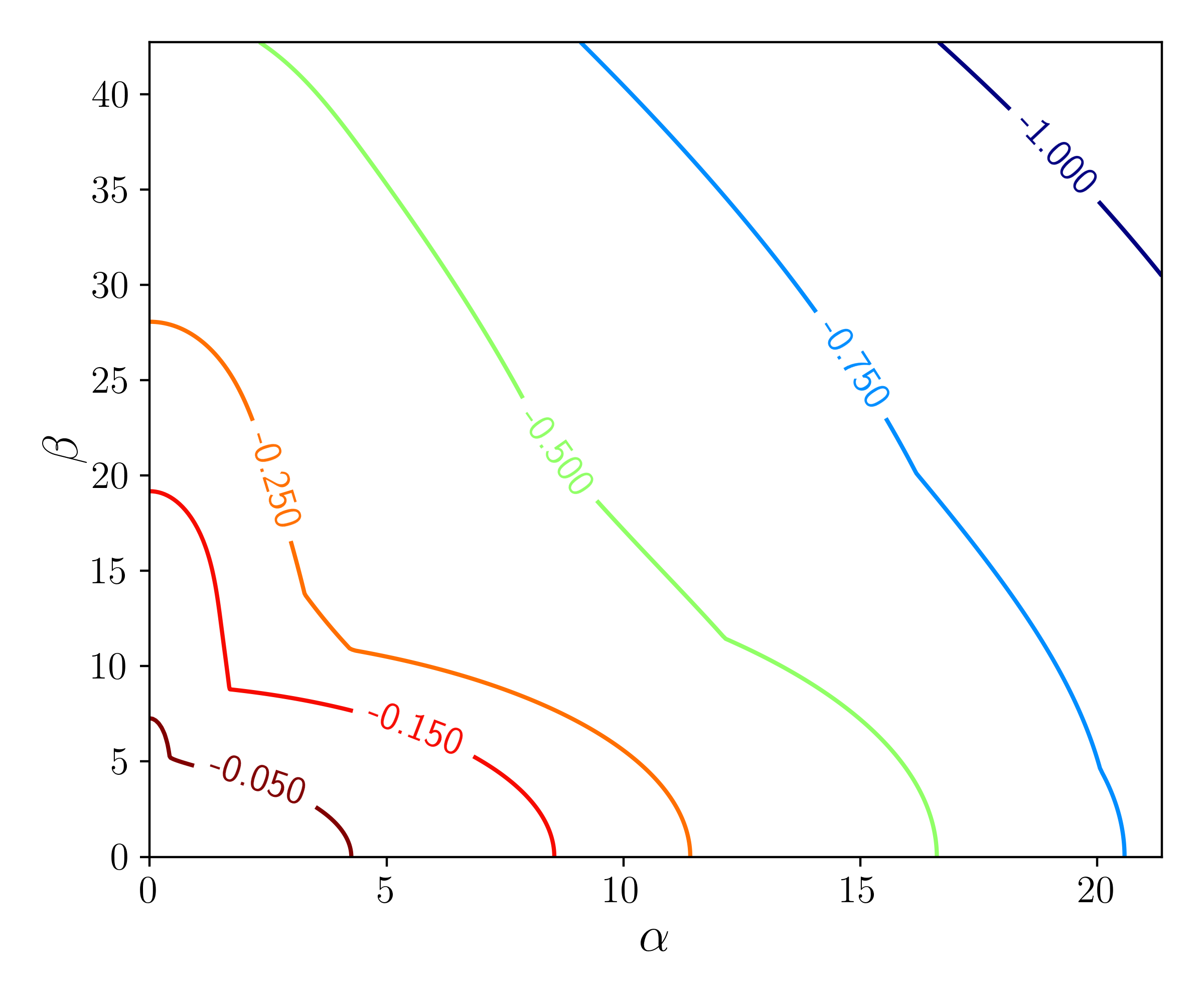}
\includegraphics[width=0.46\columnwidth]{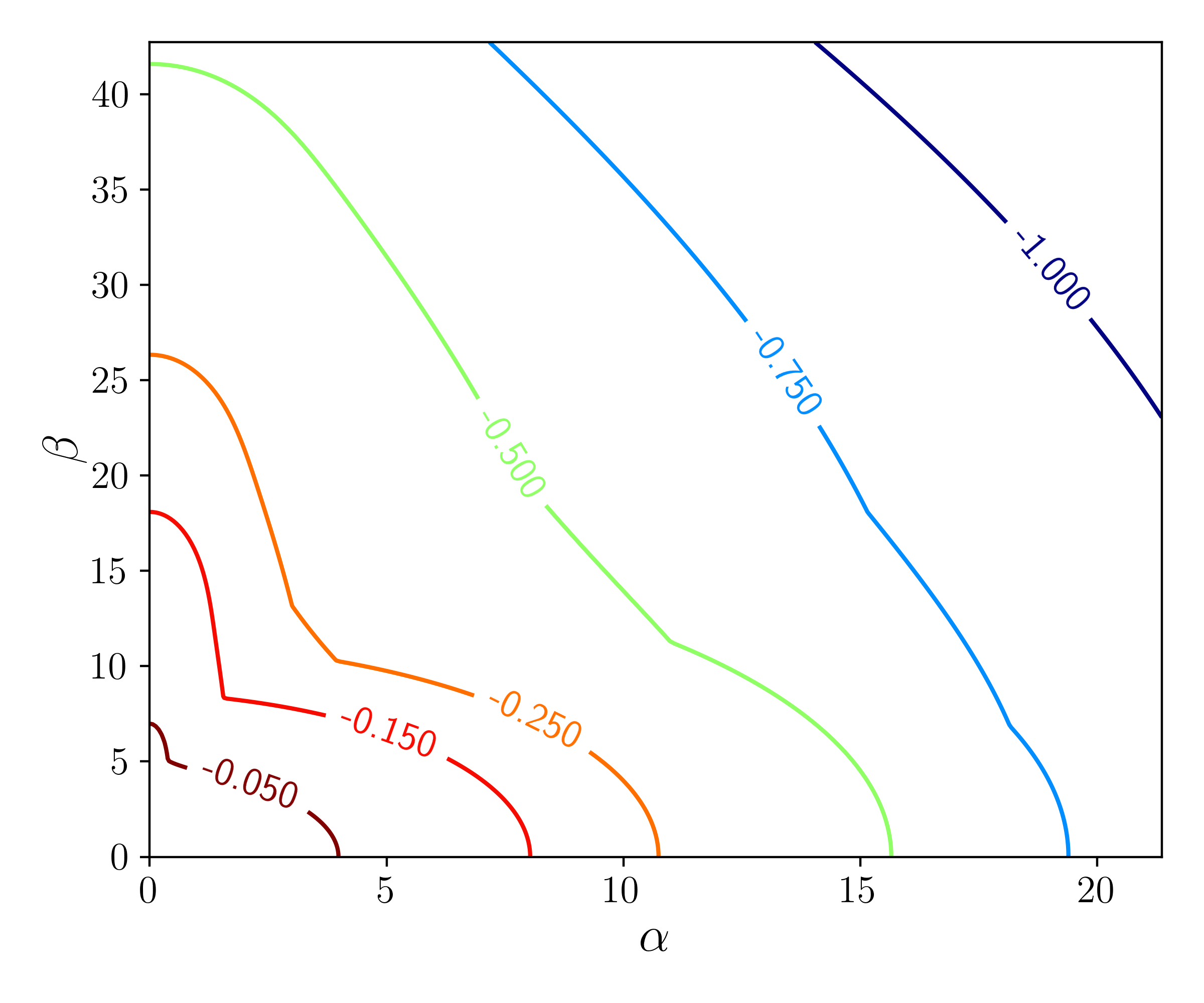}
\end{center}
\vspace{-5mm}
\caption{Isocontours of growth rate $\omega_i(\alpha,\beta)$. $Re_\tau=110$ (left) and $Re_\tau=96$ (right). }
\label{fig:ab}
\vspace{-0mm}
\end{figure}

We have explored a large range of values for the wave-numbers $(\alpha,\beta)$. Figure~\ref{fig:ab} displays the isolevels of the growth rate in the  $(\alpha,\beta)$ plane for the same two values of $Re_{\tau}$. In both cases $(\alpha,\beta)=(0,0)$ is the least stable mode. This points out the lack of selection of an oblique structure under linear processes. 

\begin{figure}
\begin{center}
\includegraphics[width=0.46\columnwidth]{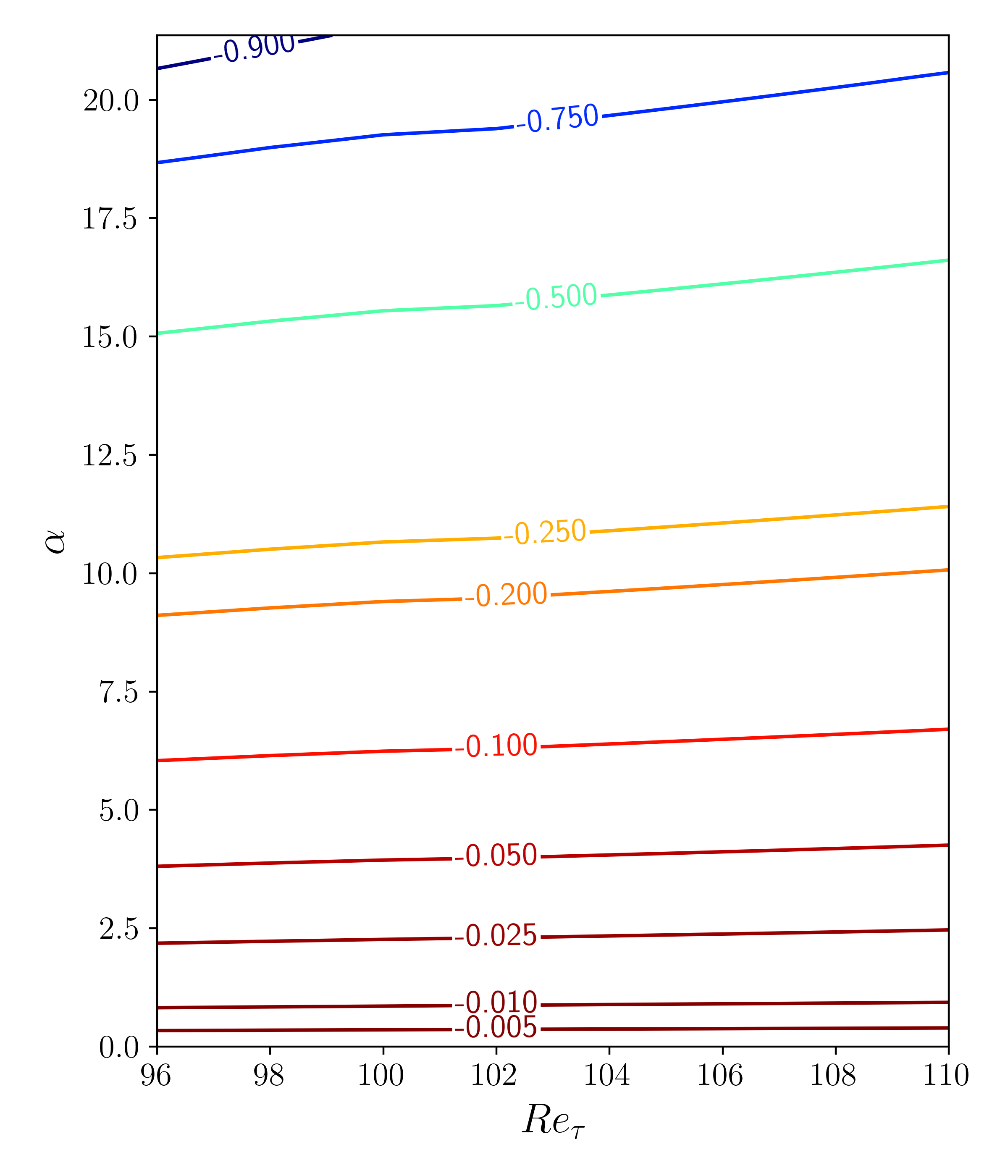}
\end{center}
\vspace{-5mm}
\caption{Stability in DNS. Isocontours of growth rate in the $(\alpha,Re_\tau)$ plane for $\beta=0$.}
\label{fig:a}
\vspace{-5mm}
\end{figure}

Finally, we explore the dependence of the growth rate on $Re_{\tau}$. As shown in Figure~\ref{fig:a} for $\beta=0$ and in Table \ref{tab-mesh}, the decay rate of the least stable mode is weakly affected by the value $Re_{\tau}$ in the range of values investigated. This result suggests that the stable eigenvalues in question are robust, and that minute modifications of other parameters (such as spatial resolution, averaging time etc...) will not have a strong influence on the conclusions.

\begin{table}[]
    \centering
    \begin{tabular}{c|ccc}
        {$Re_{\tau}$} & DNS & Cess & $A=0$  \\ \hline
         110 &  -0.00115673 & -0.00122346 & -0.00040783 \\
         102 & -0.00124783 & -0.00133929 & -0.00047432\\
         100 & -0.00127355 & -0.00137158 & -0.00049348 \\
         98 & -0.00129956 & -0.00140540 & -0.00051383\\
         96 & -0.00132725 & -0.00144086 & -0.00053546        
    \end{tabular}
    \caption{Growth rate associated with the least stable eigenvalue as a function of $Re_{\tau}$ for the three closure models investigated. 
    }
    \label{tab-mesh}
\end{table}

\begin{figure}
\begin{center}
\includegraphics[width=0.7\columnwidth]{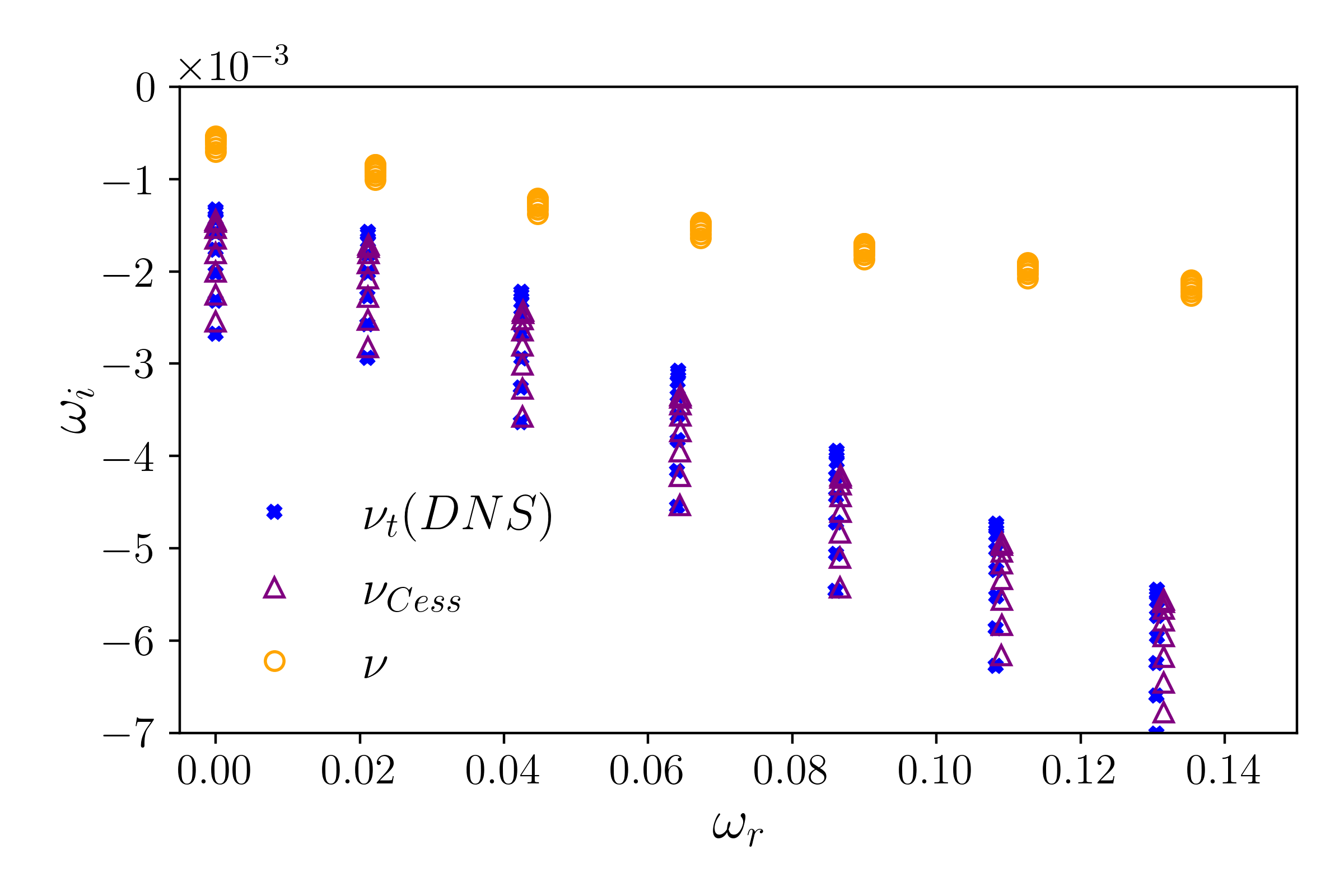}
\end{center}
\vspace{-5mm}
\caption{Eigenvalues in the complex $\omega$-plane for $Re_\tau=96$. The values of the wavenumbers are varied over the range $\alpha \le$ 0.4, $\beta \le$0.9.}
\label{fig:spectrum}
\vspace{-5mm}
\end{figure}

To conclude this analysis, we compare eigenspectra obtained from different closures (Figure~\ref{fig:spectrum}). A least stable branch of eigenvalues emerges in the case $\nu_t=0$, without any counterpart once Reynolds stresses are included, whether based on the Cess formula or on the DNS data : Reynolds stresses stabilize the homogeneous mean turbulent flow $U^*$.

\section{Discussion and outlooks}
The dynamical origin of laminar-turbulent patterns and localised turbulent bands in wall-bounded shear flows has puzzled researchers for decades. The formation of such structures can be either investigated as a finite-amplitude instability of the laminar flow, as done in several recent studies \cite{tao2018extended,xiao2020growth,parente2022linear} or directly as a linear instability of the homogeneous turbulent flow \cite{Prigent2002large}. In the case of pressure-driven channel flow, recent work has concluded that laminar-turbulent patterns emerge from homogeneous turbulent flow below a Reynolds number $Re_{\tau}$ of about 95 via a modulational instability \cite{kashyap2022linear}. In the present numerical study we have focused on the channel case and investigated the ability of linear stability analysis conducted around the mean flow to predict the onset of modulation. Several simple one-point closures have been employed to link the Reynolds stresses to the mean velocity field. In all cases the homogeneous mean flow, which depends only on the wall-normal variable $y$, appears as linearly stable for all values of the Reynolds number investigated.
Moreover the least stable mode has in all cases $k_x=k_z=0$, at odds with the non-zero wavector of the modulated state.
This result appears robust with respect to domain size and numerical resolution. It is moreover consistent with the predictions from the Squire theorem although this theorem does not strictly apply here because the base flow depends on $Re$.

There are several lessons to be learned from the failure of the analysis in predicting the observed instability.
First, because the linear instability of the mean flow is not better predicted when the turbulent viscosity profile is directly measured from the DNS data, we can conclude that it is not the specific choice of a turbulent viscosity model that matters most. 
It is therefore tempting to conclude that any one-point closure, or eddy viscosity model is doomed to fail at capturing the mechanism of the instability. In fact, this is precisely what our discussion about the applicability of the Squire transformation in Subsection \ref{sec:squire} suggests.


There are however alternative directions to investigate \. One direction is to look beyond the Boussinesq theory for the constitutive relation, another is to consider a more complex base flow. When performing the linear stability analysis of equation~\ref{eq:UNUT}, the turbulent viscosity is assumed to remain independent of $x$ and $z$. In other words, the fluctuations were averaged over the three homogeneous coordinates, respectively the streamwise direction $x$, the spanwise direction $z$ and the time $t$.
It is likely that such an extreme spatial coarsening of the fluctuations carried out here is responsible for the fact that the linear instability does not develop. In particular, at the moderate Reynolds numbers considered here, the so-called homogeneous turbulent flow is mainly structured around elongated coherent structures such as streamwise \emph{streaks} (see the $k_x-k_z$ energy spectra in Ref. \cite{kashyap2022linear}, which have a peaked distribution about $100\nu/u_{\tau}$). These structures are not properly captured by the mean flow which is based on $z$-averaging. In addition, the infinitely long time scale associated with the mean flow is probably not needed given that modulations can appear in DNS over a time scale of $O(100)$ time units \cite{kashyap2020flow}.

This suggests a more careful coarsening of the fluctuations, taking into account such structures as streaks into the base flow of interest for the stability study. The instability of optimally growing streaks has already been invoked in several works \cite{park2011stability,alizard2015linear} as a plausible explanation for the larger-scale streaks observed in high-$Re$ turbulent shear flows. Moreover, as is visually clear from Figure 1, streaks are typically small scale compared to the large scale modulations. It is thus only as a collective instability that the present large-scale modulations found below $Re \approx 100$ can arise. Investigating spatially sub-harmonic instabilities of streaks at the relevant parameter values, for instance using recent methods based on the Bloch wave formalism \cite{schmid2017stability,jouin2024collective}, is hence a promising pathway in the quest for predicting the onset of laminar-turbulent patterns.

\section*{Acknowledgements}

The authors thank O. Semeraro, C. Cossu and P. Morra for sharing their numerical code and useful discussions. J.-C. Robinet is also acknowledged for interesting discussions on the literature. L. Tuckerman and D. Barkley, as well as S. Gom\'e, are also acknowledged for discussing their unpublished work on plane Couette flow. Computations have been run at LISN, on the supercalculator Jean-Zay at IDRIS and on the cluster mesoPSL at the Observatoire de Paris, financed by the region Ile-de-France.\\

\bibliography{biblio}

\end{document}